\documentclass[journal]{IEEEtran}

\ifCLASSINFOpdf
\else
   \usepackage[dvips]{graphicx}
\fi
\usepackage{url}

\usepackage{graphicx}

\usepackage{amssymb}
\usepackage{latexsym}
\usepackage{color}
\usepackage{verbatim}
\usepackage{boldline}
\usepackage{multirow}
\usepackage{amsmath}
\usepackage{amsfonts}

\newcommand{\Z}{\ensuremath{\mathbb Z}}
\newcommand{\C}{\ensuremath{\mathbb C}}
\newcommand{\R}{\ensuremath{\mathbb R}}

\newcommand{\mC}{\mathcal{C}}

\newcommand{\mI}{\mathcal{I}}

\newcommand{\mK}{\mathcal{K}}

\newcommand{\mM}{\mathcal{M}}

\newcommand{\mO}{\mathcal{O}}
\newcommand{\mP}{\mathcal{P}}

\newcommand{\mS}{\mathcal{S}}

\newcommand{\mV}{\mathcal{V}}

\newcommand{\abu}{{\bf a}}

\newcommand{\hbu}{{\bf h}}

\newcommand{\sbu}{{\bf s}}
\newcommand{\pbu}{{\bf p}}

\newcommand{\vbu}{{\bf v}}

\newcommand{\xbu}{{\bf x}}
\newcommand{\ybu}{{\bf y}}
\newcommand{\wbu}{{\bf w}}

\newcommand{\Abu}{{\bf A}}

\newcommand{\Cbu}{{\bf C}}

\newcommand{\Fbu}{{\bf F}}
\newcommand{\Ibu}{{\bf I}}

\newcommand{\Mbu}{{\bf M}}

\newcommand{\Pbu}{{\bf P}}
\newcommand{\Rbu}{{\bf R}}
\newcommand{\Sbu}{{\bf S}}
\newcommand{\Gbu}{{\bf G}}

\newcommand{\Ubu}{{\bf U}}

\newcommand{\Wbu}{{\bf W}}
\newcommand{\Xbu}{{\bf X}}
\newcommand{\Ybu}{{\bf Y}}
\newcommand{\Zbu}{{\bf Z}}

\newcommand{\balp}{{\boldsymbol \alpha}}

\newtheorem{rem}{Remark}

\numberwithin{const2}{const}

\begin{document}

\title{Design of Non-Orthogonal Sequences Using a Two-Stage Genetic Algorithm for Grant-Free Massive Connectivity}

\author{Nam Yul Yu, \IEEEmembership{Senior Member,~IEEE}
\thanks{This work has been submitted to the IEEE for possible publication. 
	Copyright may be transferred without notice, after which this version may no longer be accessible.
	This work was supported by the National Research Foundation of Korea (NRF) grant funded by the Korea Government (MSIT)
	(NRF-2021R1F1A1046282).}
\thanks{The author is with the School of
of Electrical Engineering and Computer Science (EECS), Gwangju Institute of Science and Technology (GIST), Korea.
(e-mail: nyyu@gist.ac.kr).}
}  

\maketitle

\begin{abstract}
In massive machine-type communications (mMTC),
grant-free access is a key enabler for a massive number of users
to be connected to a base station with low signaling overhead and low latency.
In this paper, a two-stage genetic algorithm (GA) is proposed 
to design a new set of user-specific, non-orthogonal, unimodular sequences
for uplink grant-free access.
The first-stage GA is
to find a subsampling index set for a partial unitary matrix 
that can be approximated to an equiangular tight frame.
Then in the second-stage GA, we try to find a sequence to be masked to each column
of the partial unitary matrix, 
in order to reduce the peak-to-average power ratio of the resulting columns
for multicarrier transmission.
Finally, the masked columns of the matrix are proposed as new non-orthogonal sequences  
for uplink grant-free access.
Simulation results 
demonstrate that the non-orthogonal sequences designed by our two-stage GA
exhibit excellent performance 
for compressed sensing based joint activity detection and channel estimation
in uplink grant-free access.
Compared to algebraic design,
this GA-based design 
can present a set of good non-orthogonal sequences of arbitrary length, 
which provides more flexibility for uplink grant-free access in mMTC.
\end{abstract}

\begin{IEEEkeywords}
Compressed sensing, genetic algorithm, grant-free access, machine-type communications, non-orthogonal multiple access, 
peak-to-average power ratio. 
\end{IEEEkeywords}

\IEEEpeerreviewmaketitle

\section{Introduction}

\IEEEPARstart{M}{assive} connectivity of wireless devices
is essential for industrial, commercial, and critical applications
of massive machine-type communications (mMTC)~\cite{Kunz:MTC, Bockel:mMTC},
which provides a concrete platform for the Internet of Things (IoT).
Unlike human-type communications (HTC), 
mMTC is characterized by small size data, infrequent transmission, low cost devices,
low mobility, and so on~\cite{3gpp:22368}.
In practice,
mMTC systems need to support a massive number of devices with low control overhead, low latency,
and low power consumption for delay-sensitive and energy efficient communications.

Non-orthogonal multiple access (NOMA)~\cite{Dai:noma, Dai:survey} has received a great deal of attention 
for massive connectivity in 5G wireless systems.
In code-domain NOMA, user-specific and
non-orthogonal spreading sequences
are assigned to users for their non-orthogonal multiplexing through common resources. 
In sparse code multiple access (SCMA)~\cite{Baligh:SCMA}, 
sparse spreading sequences are assigned to users, 
where a message passing algorithm (MPA)~\cite{Zhang:MPA} and a list sphere decoding based MPA decoder~\cite{Wei:SCMA}
can be deployed for reliable multiuser detection with low complexity.  
Complex-valued spreading sequences are employed for 
multi-user shared access (MUSA)~\cite{Yuan:MUSA}, 
where the successive interference cancellation (SIC) can be performed for multiuser detection. 
Also, pattern division multiple access (PDMA)~\cite{Kang:PDMA} attempted
to enable massive connectivity with low complexity through an efficient pattern matrix design 
and a recursive approach of multiuser detection~\cite{Jamali:PDMA}.
For a survey on existing works of code-domain NOMA, readers are referred to~\cite{Dai:survey}.
Recently, the state-of-the-art technique of deep learning~\cite{Bengjo:DL} has been applied
for multiuser detection in uplink code-domain NOMA systems~\cite{Ye:deep}$-$\cite{Siva:dnn}.

Grant-free access is of tremendous interest to connect a massive number of users to mMTC systems
with low latency and low signaling overhead~\cite{Cirik:toward}.
In uplink grant-free access, 
active users send their data with no access-grant procedure.
Then, a base station (BS) receiver has to identify active users with no aid of a grant procedure
and detect each active user's data from the superimposed signal.
The principle of 
compressed sensing (CS)~\cite{Eldar:CS} can be applied for multiuser detection in uplink grant-free access,  
exploiting the sparse activity that many users are present in a cell, 
but only a few of them are active at a time.
Many research articles~\cite{Abebe:iter}$-$\cite{Schober:meet} demonstrated that 
a CS-based detector can be successfully deployed at BS  
for joint activity detection, channel estimation, and/or data detection in uplink grant-free access.

For non-orthogonal and grant-free access, 
it is crucial to design a set of non-orthogonal sequences with low correlation,
constructively or algorithmically, which 
ultimately guarantees reliable 
CS-based detection at BS.
Moreover, if the transmitted signals of active users are spread 
onto multiple subcarriers,
the high peak-to-average power ratio (PAPR) 
will cause signal distortion
deteriorating all potential benefits of multicarrier communications~\cite{Litsyn:peak, No:papr}.
Various reduction techniques~\cite{No:papr} have been proposed for mitigating the PAPR
of multicarrier transmitted signals. 
Recently, efforts have been made to reduce the PAPR of uplink multicarrier signals in SCMA~\cite{Yani:ra, Mherich:golden}.
In summary, we need to design a set of
non-orthogonal sequences with low correlation and low PAPR properties,
which ensures reliable and power efficient uplink grant-free access. 

In literature,
many constructive designs have been presented 
to provide a variety of pilot or spreading sequences for 
multiple access.
In~\cite{Yang:qos}, quasi-orthogonal sequences have been introduced
to increase the system capacity of CDMA.
Random sequences with the Gaussian distributed elements have been used in~\cite{Liu:mimo}$-$\cite{Jiang:noma}
to theoretically guarantee reliable CS-based detection 
for uplink access.
Also, the works of~\cite{Wang:struct}$-$\cite{Du:block} 
used pseudo-random noise sequences 
for CS-based detection in uplink grant-free NOMA.
In multicarrier communications,
Golay complementary sequences and sets~\cite{Golay:series}$-$\cite{Paterson:gen} 
can be employed to provide theoretically bounded low PAPR.
In~\cite{Liu:FBMC}, Golay complementary sequences have been also applied for 
low PAPR preambles in the filter-bank multicarrier (FBMC) modulation.
Binary~\cite{Yu:binary} and non-binary~\cite{Yu:non} Golay spreading sequences have been employed for low PAPR in uplink grant-free NOMA.
Other complementary sequences have been studied in~\cite{Liu:comp}$-$\cite{Wu:Z}
for PAPR reduction. 
Zadoff-Chu (ZC) sequences~\cite{Chu:ZC}, also known as constant amplitude and zero autocorrelation (CAZAC) sequences,
have been adopted as preambles for random access
in 3GPP-LTE~\cite{3gpp:36.211},
providing low PAPR for multicarrier transmission.

Noting that a sensing matrix of CS is a collection of non-orthogonal column sequences,
we can find many algorithmic approaches 
for good sequences from the efforts of optimizing the sensing matrix.
Elad~\cite{Elad:opt} launched an algorithmic design for a sensing matrix
by minimizing the average measure of the coherence iteratively.
In~\cite{Xu:opt}$-$\cite{Zhang:opt}, several algorithms have been proposed for
optimizing a sensing matrix, 
where each one attempts to approximate 
its Gram matrix to that of an equiangular tight frame (ETF)~\cite{Kov:frames}. 
In~\cite{Chen:utf}, Chen~\emph{et al.} demonstrated that
a unit-norm tight frame is a closest design of a nearly orthogonal matrix. 
Other algorithms can be found in~\cite{Duarte:learn} and \cite{Lu:dir}. 
From these efforts, each sensing matrix optimized algorithmically can offer a set of 
non-orthogonal sequences 
for reliable CS-based detection.
In~\cite{Chun:DL} and \cite{Kim:DL}, 
deep learning (DL) techniques have been also applied for pilot or spreading sequence design. 
In general, the non-orthogonal sequences obtained by algorithmic and DL-based designs 
can take arbitrary elements with no structure,
which may not be suitable for cost efficient implementation in mMTC devices.

Recently, the genetic algorithm (GA)~\cite{Holland:gen} has been applied 
for sensing matrix optimization in specific applications,
e.g., reducing the complexity of 
radar imaging~\cite{Chen:ISAR}, allocating an optimized pilot pattern for channel estimation~\cite{Nie:maga}, 
and maximizing the energy efficiency of wireless sensor networks (WSN)~\cite{Leven:moga}.
In particular, 
GA has been used
to find subsampling patterns to optimize partial Fourier matrices
with specific parameters~\cite{Chen:ISAR, Nie:maga}.
This GA-based optimization motivates us to scrutinize the effectiveness of GA 
for optimizing a sensing matrix, which can ultimately present a set of 
non-orthogonal sequences for uplink grant-free access.

In this paper, 
we propose a two-stage genetic algorithm (GA) to design a new set of non-orthogonal sequences\footnote{
The resulting sequences from our design can be used as
spreading, pilot, or signature sequences, depending on specific access schemes.} 
for uplink grant-free access, 
where each sequence has unimodular and complex-valued elements of finite phase for
cost efficient implementation in an mMTC device.
Each stage of GA makes an evolutionary approach to reach an optimized result by transforming
and improving the intermediate outcomes. 
The first-stage GA
is to find a subsampling index set to optimize\footnote{In this paper, 
`\emph{optimize}' does not mean to find a global optimum, since GA may converge to local optima.} 
a partial unitary matrix by
approximating it to an ETF,
where the evolutionary approach tries to minimize  
the average distance between the inner product of its column pair and the 
Welch bound equality~\cite{Welch:low}.
Then, the second-stage GA tries to find a sequence to be commonly masked 
to each column of the partial unitary matrix from the first-stage,
in order to reduce the PAPR of the resulting columns.
Note that masking each column with a common sequence does not change
the inner products and their distribution among the resulting column pairs.
Finally, the masked columns of the partial unitary matrix
are proposed as new non-orthogonal sequences with low correlation and low PAPR properties,
which can be uniquely assigned to users  
for uplink grant-free access.

Through simulations, the phase transition diagrams
reveal that the
partial Fourier and ZC-based matrices optimized by our first-stage GA guarantee more reliable CS reconstruction
than the randomly subsampled counterparts, respectively, 
over a wide range of compression and sparsity ratios.
In addition, it turns out that the second-stage GA is effective to enhance
the PAPR properties of the resulting sequences, where 
the PAPR distributions appear to be acceptable for multicarrier transmission.
In uplink grant-free access,
we show that the performance of the Fourier- and ZC-based sequences from our two-stage GA
is superior to that of random sequences, while comparable to that of ZC sequences of prime length,
for CS-based joint activity detection and channel estimation
Compared to algebraic design, we confirm that this GA-based design can present a new set of
non-orthogonal sequences of arbitrary length,
exhibiting acceptable PAPR distribution and 
guaranteeing reliable CS-based detection,
which can be more suitable for 
grant-free massive connectivity.

This paper is organized as follows.
Section II describes a system model of uplink grant-free access under consideration,
where a CS problem is formulated for joint activity detection and channel estimation.
Section III outlines a framework for non-orthogonal sequence design
using a two-stage GA. In each stage, we formulate the design goal by an optimization problem.
Section IV describes the evolution steps of each stage GA 
along with the cost function for the optimization problem.
Algorithms 1 and 2 summarize the two-stage GA.
Section V presents simulation results to demonstrate the effectiveness of each stage GA.
In addition, we evaluate the performance of the proposed sequences, comparing to other conventional ones,
in CS-based joint activity detection and channel estimation. 
Finally, concluding remarks will be given in Section IV.

\emph{Notations}: 
Throughout this paper, $\Z_N = \{0, \cdots, N-1\}$.
A matrix (or a vector) is represented by
a bold-face upper (or a lower) case letter.
$\Xbu^T$ denotes the transpose of a matrix $\Xbu$, 
while 
$\Xbu^*$ is its conjugate transpose. 
The identity matrix is denoted by $\Ibu$, where the dimension is determined in the context.
$ {\rm abs} (\Xbu) = [ | X_{i, j} | ]$
denotes a matrix taking the magnitude of each element of $\Xbu =  [X_{i,j}]$.
For a vector $\hbu $, 
$\hbu_{\mS}$ is its subvector, indexed by an index set $\mS$, and
${\rm diag} (\hbu)$ is a diagonal matrix whose diagonal entries are from $\hbu$.
The inner product of vectors $\xbu$ and $\ybu$ is denoted by $\langle \xbu, \ybu \rangle$.
The $l_2$-norm of a vector $\xbu  = (x_1, \cdots, x_N)$ is denoted by
$ || \xbu ||_{2} = \sqrt{ \sum_{k=1} ^{N} |x_k|^2 } $.
The Frobenius norm of a matrix $\Xbu = [X_{i,j}]$ is denoted by 
$\| \Xbu \|_F = \sqrt{\sum_{i, j} \left| X_{i,j} \right| ^2 }$.
Finally, $\hbu \sim \mathcal{CN} (\mathbf{m}, \mathbf{\Sigma})$ is a circularly symmetric complex Gaussian random vector
with mean $\bf m$ and covariance $\mathbf{\Sigma}$.

\section{System Model}
\begin{figure}
	\centering
	\includegraphics[width=0.48\textwidth, angle=0]{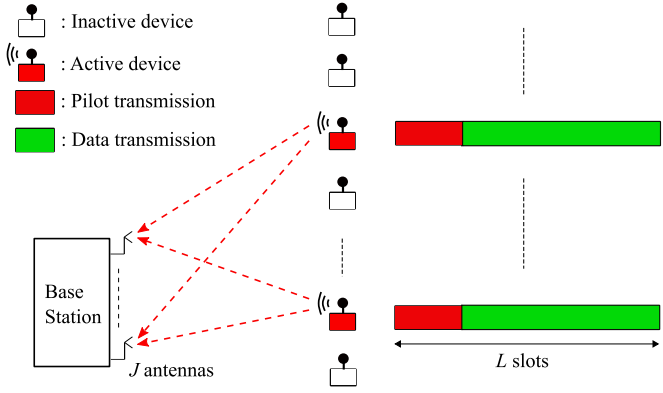}
	\caption{Two-phase grant-free access scheme with multiple receiver antennas.}
	\label{fig:system}
\end{figure}

In this paper, we consider a two-phase grant-free access scheme~\cite{Liu:mimo, Liu:massive}
for a single-cell massive connectivity. 
In an mMTC cell, a base station (BS) receiver equipped with $J$ antennas
accommodates total $N$ devices each of which transmits with a single antenna.
For a fully grant-free access, 
we assume that devices are \emph{static}\footnote{
	If devices are moving from cell to cell, 
	it is hard to guarantee unique sequence assignment in a fully grant-free manner
	for all devices in a cell, and 
	some coordination may be required to assign unique sequences to devices.}
in a cell and thus BS accommodates a fixed set of devices
having their own user-specific sequences. 
In the first phase, each active device transmits its sequence as a dedicated pilot,
and the BS receiver then tries to identify active devices and estimate their channel profiles 
from the superimposed pilots.
Data can be directly transmitted in the second-phase from active devices with no grant from BS.
In this two-phase scheme, 
we assume that the channels and the device activity remain unchanged during $L$ slots for pilot 
and data transmissions.
Figure~\ref{fig:system} illustrates this system model.

With sparse activity, 
each device is assumed to be active with probability $p_a$ in an i.i.d. manner,
where active devices are synchronized.
In an access time, 
an activity indicator vector can be defined by $\balp = (\alpha_1, \cdots, \alpha_N)^T$ with
\begin{equation*}\label{eq:act}
	\alpha_n = \left\{ \begin{array}{ll} 1, & \mbox{ if device } n \mbox{ is active}, \\
	0, & \mbox{ otherwise}, \end{array} \right.
\end{equation*}
where $\mS = \{ n \mid \alpha_n = 1 \}$ is a set of active devices and
the number of active devices is $|\mS| = \sum_{n=1} ^N \alpha_n = K \ll N$.

When device $n$ is active, 
it transmits its unique pilot
sequence $\sbu_n = (s_{1,n}, \cdots, s_{M,n})^T $ 
over $M$ subcarriers for grant-free access, where $M < N$.
We consider a flat Rayleigh fading channel, 
where the channel gain remains unchanged during the coherence time interval of $L$ slots.
Let $\hbu_n = \left(h_n ^{(1)}, \cdots, h_n ^{(J)} \right)^T$, $ 1 \leq n \leq N$,
be a channel vector from device $n$, 
where $h_n ^{(t)}$ is the channel gain between device $n$ and BS receiver antenna $t$.
Assuming that the path loss and shadowing effects are known and can be removed by BS,
we have $\hbu_n \sim \mathcal{CN}(\bf{0}, \Ibu) $. 
Then, the received signal at antenna $t$ can be represented 
by
\begin{equation}\label{eq:y}
	\ybu^{(t)} = \sum_{n=1} ^N \alpha_n  h_n ^{(t)} \sbu_n  + \wbu ^{(t)}
	=  \Sbu \xbu ^{(t)} + \wbu ^{(t)}, 
\end{equation}
where $\xbu ^{(t)} = \left(\alpha_1 h_1 ^{(t)}, \cdots, \alpha_N h_N ^{(t)} \right)^T$
for $ 1 \leq t \leq J$. 
In~\eqref{eq:y},
$\Sbu = [\sbu_1, \cdots, \sbu_N] \in \C^{M \times N} $ is a matrix of pilot sequences,
and $\wbu^{(t)} \sim \mathcal{CN}(\textbf{0}, \sigma_n ^2 \Ibu) $ is the complex Gaussian noise vector
at antenna $t$.

Collecting the $J$ received signals of~\eqref{eq:y}, 
we have a multiple measurement vector (MMV) 
model of
\begin{equation}\label{eq:mmv}
\Ybu = \Sbu \Xbu + \Wbu,
\end{equation}
where $\Ybu = \left[\ybu^{(1)}, \cdots, \ybu^{(J)} \right]$,
$\Xbu = \left[\xbu^{(1)}, \cdots, \xbu^{(J)} \right]$,
and $\Wbu = \left[\wbu^{(1)}, \cdots, \wbu^{(J)} \right]$, respectively.
Due to the activity indicator $\balp$, 
it is clear that $\Xbu$ has the row-wise sparsity with $K$ nonzero 
and $N-K$ zero rows.
Then, BS can apply
a joint sparse recovery algorithm to solve the MMV problem of~\eqref{eq:mmv},
in order to detect the activity indicator $\balp$ and estimate the channel vector $\hbu_n$ for $n \in \mS$.
If the nonzero rows of $\Xbu$ are estimated,
the row indices mean a detected index set of active devices, denoted by $\widehat{\mS}$, 
while the coefficients of each nonzero row give 
an estimated channel vector $\widehat{\hbu}_n$ for $n \in \widehat{\mS}$. 
The CS-based joint active user detection (AUD) and channel estimation (CE)
complete the first phase of uplink grant-free access.
In the second phase, the BS receiver detects data from active devices
with the knowledge of device identity and channel profiles obtained from the first phase~\cite{Liu:mimo, Liu:massive}.
In this paper, we restrict our attention to joint AUD and CE in the first phase
via joint sparse recovery under the CS MMV model.

\begin{rem}\label{rm:one}
A CS MMV model can also be applied for \emph{one-shot} detection
in uplink grant-free NOMA~\cite{Du:joint, Yu:binary}.
In this system, each active device transmits its unique spreading sequence of length $M$,
spread onto $M$ subcarriers,  
carrying its pilot and data over $J$ time slots.
Assuming that the channels and the device activity remain unchanged, 
the received signals over $J$ slots are also modeled by \eqref{eq:mmv}.
A BS receiver equipped with a single antenna
then conducts CS-based joint activity detection, channel estimation, and data detection,
by solving the MMV problem of \eqref{eq:mmv}.
Readers are referred to~\cite{Du:joint} and \cite{Yu:binary} for more details.
\end{rem}

\section{Framework for Sequence Design}
The goal of this paper is to present a set of non-orthogonal sequences
for grant-free massive connectivity.
In CS MMV model, the sequence set forms the matrix $\Sbu$ in \eqref{eq:mmv}, where
the problem of sequence design
boils down to designing a sensing matrix for reliable CS-based detection.
This section outlines a framework for sensing matrix design using the genetic algorithm (GA),
which ultimately provides a set of good non-orthogonal sequences for uplink grant-free access.

\subsection{Partial Unitary Matrices}

Compressed sensing (CS)~\cite{Eldar:CS} 
is to reconstruct an $N$-dimensional sparse signal $\xbu$
from its underdetermined $M$-dimensional measurement $\ybu = \Abu \xbu$, 
where $M < N$.
The signal $\xbu $ is called \emph{$K$-sparse}
if it has at most $K$ nonzero elements,
where $K \ll N$. 
In CS techniques, it is essential to design a good $M \times N$ sensing matrix $\Abu$,
in order to guarantee reliable reconstruction of sparse signals.
Taking some rows out of a unitary matrix
is a well known operation to obtain a \emph{partial} unitary matrix~\cite{Duarte:struct} 
that 
enjoys practical benefits as well as theoretical CS recovery guarantee.
A partial unitary matrix is formulated by
\begin{equation}\label{eq:S-single}
	\Abu  =\frac{1}{\sqrt{M}} \mathbf{R}_{\Omega}\Ubu \triangleq \Ubu_\Omega,
\end{equation}
where $\Ubu$ is an $N \times N$ unitary matrix of $\Ubu \Ubu^* = \Ubu^* \Ubu = N \Ibu$.
In~\eqref{eq:S-single}, $\mathbf{R}_\Omega$ is a subsampling operator selecting 
$M$ rows out of $N$ ones whose indices are specified by $\Omega \subset \{1, \cdots, N\}$, where $|\Omega| = M$. 
If the indices of $\Omega$ are selected randomly, 
$\Abu = \Ubu_\Omega$ 
guarantees reliable CS reconstruction 
theoretically with high probability,
provided that $M \geq \mO ( K \log^4 N)$~\cite{Rud:sparse}.
In practice, a partial unitary matrix allows fast and efficient measurement and reconstruction for CS,
thanks to the fast unitary transform, e.g., fast Fourier or Hadamard transform.

To design unimodular sequences, 
we begin with a unitary matrix $\Ubu$ whose elements take the magnitude of $1$.
Then, the first stage of sequence design 
attempts to find a subsampling index set $\Omega$
to optimize a partial unitary matrix $\Ubu_\Omega$ for reliable CS reconstruction.

\subsection{PAPR Reduction}
In system model of Section II, 
if the sequence $\sbu_n $ 
is transmitted through $M$ subcarriers,
the peak-to-average power ratio (PAPR) of its OFDM signal is determined by~\cite{Litsyn:peak}
\begin{equation}\label{eq:papr}
	{\rm PAPR} (\sbu_n)= \max_{t \in [0, 1) } 
	\frac{ \left| \sum_{i=1} ^{M} s_{m, n} e^{j 2 \pi (m-1) t} \right|^2 }{ M} ,
\end{equation}
where $j=\sqrt{-1}$.
In~\eqref{eq:papr}, we assumed that $\sbu_n $ is unimodular, i.e., $|s_{m, n}| =1$, for $m=1, \cdots, M$. 

Given a partial unitary matrix $\Ubu_\Omega$, 
we try to reduce the PAPR of
the column sequences\footnote{
If $\Ubu$ has a column of all ones, like the Fourier or Hadamard matrix,
the maximum PAPR of the column sequences of its partial unitary matrix $\Ubu_\Omega$ has the highest value of $M$,
regardless of $\Omega$.}.
For PAPR reduction, we apply
a unimodular and complex-valued sequence $\vbu = (v_1, \cdots, v_{M})^T$   
as a common mask to each column of $\Ubu_\Omega$, i.e.,
\begin{equation}\label{eq:mask}
\Ubu_{\Omega, \vbu}  
= {\rm diag} (\vbu) \cdot \Ubu_\Omega.
\end{equation}
In~\eqref{eq:mask}, we use
a modulated $q$-ary sequence for $\vbu$, i.e., $v_m = e^{j \frac{2 \pi a_m}{q}}$,
where $a_m \in \Z_q$ for $m=1, \cdots, M$.
Then, it is clear that the inner product of a column pair in $\Ubu_{\Omega, \vbu}$
is identical to that of the corresponding pair in $\Ubu_\Omega$, since
$\Ubu_{\Omega, \vbu} ^* \Ubu_{\Omega, \vbu} = \Ubu_\Omega ^* \Ubu_\Omega$,
which suggests that the new matrix $\Ubu_{\Omega, \vbu}$ may exhibit 
the same performance of CS-based detection as the matrix $\Ubu_\Omega$.

In the second design stage, we search for a masking sequence $\vbu$
that allows the column sequences of $\Ubu_{\Omega, \vbu}$ to have a desired PAPR property,
maintaining the performance of reliable CS reconstruction from the first design stage.


\subsection{Genetic Algorithm}
The genetic algorithm (GA) is an evolutionary technique
to solve an optimization problem that is computationally intractable~\cite{Holland:gen}.
Inspired by the evolutionary mechanism in nature,
GA transforms and evolves \emph{chromosomes} through crossover, mutation, selection, population updates, and so on.
Through a sufficient number of generations,
GA converges to a fittest chromosome, which can be a solution to the optimization problem. 
Thanks to the fast convergence to local optima,
GA has attracted much attention in machine learning and data mining~\cite{Freitas:survey}$-$\cite{Viv:intel}.
Recently, GA has expanded its application to other areas, e.g., 
channel coding~\cite{Hebbes:turbo}$-$\cite{brink:polar}, spreading code design~\cite{Dam:seq, Nat:evol},
CS recovery~\cite{Conde:sparse}$-$\cite{Erkoc:evol}
and matrix optimization~\cite{Chen:ISAR}$-$\cite{Leven:moga}, etc.

In Section III.A and III.B, 
we introduced two design stages to obtain a set of good sequences.
At each stage, the design goal can be specified by
an optimization problem that needs to be solved by GA. 
Given a unitary matrix $\Ubu$, 
the objective of the first design stage is to find a fittest chromosome 
or subsampling index set $\Omega$ in~\eqref{eq:S-single}. 
The optimization problem 
for this objective can be formulated by
\begin{equation}\label{eq:opt}
	\Omega_{\rm opt} = \underset{\Omega \subset \{1, \cdots, N\}, |\Omega| = M}{\mbox{argmin}} \ f_1 (\Ubu_\Omega), 
\end{equation} 
where $f_1 (\Ubu_\Omega)$ is a cost function of the first-stage optimization.
The cost function needs to be a good metric 
that reflects the performance of CS reconstruction
with the partial unitary matrix $\Ubu_{\Omega}$. 
The first-stage GA tries to minimize the cost function $f_1 (\Ubu_\Omega)$ through evolution steps,
in order to find a solution to~\eqref{eq:opt}.

When the first-stage GA is completed, 
the partial unitary matrix $\Ubu_{\Omega_{\rm opt}}$ with an optimized subsampling index set $\Omega_{\rm opt}$ 
is available for the second design stage. 
Given $\Omega_{\rm opt}$, 
the second-stage GA tries to find a fittest chromosome or unimodular masking sequence $\vbu$ of length $M$,
which is a solution to another optimization problem of
\begin{equation}\label{eq:opt2}
	\vbu_{\rm opt} = \underset{\vbu \in \mV_{q, M}}{\mbox{argmin}} \ f_2 (\Ubu_{\Omega_{\rm opt}, \vbu}) ,
\end{equation}
where $\mV_{q, M}$ is a set of all modulated $q$-ary sequences of length $M$.
In~\eqref{eq:opt2}, $f_2 (\Ubu_{\Omega_{\rm opt}, \vbu})$ is a cost function of 
the second-stage optimization, which
should be a metric for the PAPR property of 
the column sequences of $\Ubu_{\Omega_{\rm opt}, \vbu} $.
The second-stage GA tries to enhance the PAPR property with $\vbu_{\rm opt}$,
maintaining the performance of $\Ubu_{\Omega_{\rm opt}}$ for reliable CS reconstruction.
 
Finally, if $\Omega_{\rm opt}$ and $\vbu_{\rm opt}$ are found by the two-stage GA,
we obtain the matrix $\Sbu = \Ubu_{\Omega_{\rm opt}, \vbu_{\rm opt}}$ in~\eqref{eq:mmv}, 
where the column sequences are proposed as non-orthogonal sequences with low PAPR  
for reliable CS-based detection in uplink grant-free access.
In next section, our two-stage GA will be described 
with the details 
to find $\Omega_{\rm opt}$ and $\vbu_{\rm opt}$, respectively.


\section{Two-Stage Genetic Algorithm}
In this section, we describe the evolutionary steps of our two-stage GA. 
The first-stage GA is to find an optimized subsampling index set $\Omega_{\rm opt} \subset \{1, \cdots, N\}$ 
with $|\Omega_{\rm opt}| = M$, where $M$ and $N$ are fixed.
Given $\Omega_{\rm opt}$, 
the second-stage GA then attempts to find an optimized masking sequence $\vbu_{\rm opt} \in \mV_{q, M}$,
where $q = N$.

\subsection{Stage 1: Subsampling Optimization}

\subsubsection{Initialization}
A population $\mP_1$ is defined by a collection of $T_1$ subsampling index sets, i.e.,
$\mP_1 = (\Omega_1, \Omega_2, \cdots, \Omega_{T_1})$,
where $\Omega_t \subset \{1, \cdots, N\}$ with $|\Omega_t| = M$ for $t = 1, \cdots, T_1$.
Initially, the indices of $\Omega_t$ are selected randomly. 

\subsubsection{Cost Function}
In the first-stage, we propose the cost function for an index set $\Omega_t$ by
\begin{equation}\label{eq:cost}
	f_1(\Ubu_{\Omega_t}) = \frac{1}{\sqrt{N(N-1)}} \left \| {\rm abs} \left(\Ubu_{\Omega_t} ^* \Ubu_{\Omega_t} \right) - \Gbu_W \right \|_F,  
\end{equation}
where $\Gbu_W \in \R^{N \times N}$ is a matrix with diagonal entries of $1$ and off-diagonal entries 
of $\sqrt{\frac{N-M}{M(N-1)}}$.
Intuitively, the cost function of \eqref{eq:cost} represents
the average (rms-sense) distance between the inner product of a column pair of $\Ubu_{\Omega_t}$ 
and the Welch bound equality (WBE)~\cite{Welch:low}.
Attempting to minimize the cost function, 
the first-stage GA 
makes the inner product of a column pair of $\Ubu_{\Omega_t}$ closer to the WBE,
which approximates the resulting matrix to an equiangular tight frame (ETF)~\cite{Kov:frames}.
The target matrix $\Gbu_W$ is similar to, but not the same as the one in  
the convex set (e.g. (12) in~\cite{Vahid:grad} and (14) in~\cite{Li:opt}) for 
optimizing CS matrices.
As remarked by \cite{Li:opt},
it is more reasonable to measure a distance from $\Gbu_W$ in~\eqref{eq:cost}, rather than $\Ibu$, 
which has been confirmed by the optimization of~\cite{Vahid:grad}. 

\subsubsection{Crossover}
In population $\mP_1$,
let us 
consider a pair of index sets $\Omega_{t_1}$ and $\Omega_{t_2}$, $1 \leq t_1 \neq t_2 \leq T_1$, where
we assume
$f_1 \left(\Ubu_{\Omega_{t_1}} \right) < f_1 \left(\Ubu_{\Omega_{t_2}} \right)$. Then,
$d_1 = \lceil \beta_1 \cdot M \rceil$ and 
$d_2 = M - d_1$ indices are randomly selected from $\Omega_{t_1}$ and $\Omega_{t_2}$, respectively,
where $\beta_1 > 0.5$.
Finally, the selected indices, which should be all distinct, 
are combined to generate a new index set through \emph{crossover}. 
In other words, we create a new subsampling index set by combining parents,
where a parent index set with a lower cost function
is more involved in creating its offspring.
We apply the crossover for every pair of parent index sets from $\mP_1$, 
which yields a new population $\mC_1$ of size $ \binom{T_1}{2} = \frac{T_1(T_1-1)}{2}$ at each evolution step.

\subsubsection{Mutation}
In nature,
parts of a chromosome can be mutated in a generation,
which provides diversity for evolution.
In the first-stage GA, 
$\mu_1$ indices are randomly selected from each index set 
in $\mP_1$,
which is then replaced by new (random) ones through \emph{mutation}.
At each evolution step, 
we apply the mutation to all index sets in $\mP_1$, 
which yields a new population $\mM_1$ of size $T_1$.

\subsubsection{Population Update}
Through crossover and mutation,
we have a new, intermediate population $\mI_1 = \mP_1 + \mC_1 + \mM_1$,
where the size\footnote{
Identical chromosomes in $\mI_1$ (or $\mI_2$), if any, are treated as separate ones.} 
is $ |\mI_1| = T_1 + \binom{T_1}{2} + T_1 =  \frac{T_1(T_1+3)}{2}$.
From $\mI_1$, we select the $T_1$ index sets 
with the $T_1$ lowest cost functions of \eqref{eq:cost}.
The population $\mP_1$ is then updated by the $T_1$ fittest index sets
at each evolution step. 

\subsubsection{Iteration and Selection}
In the first-stage GA, 
the evolution steps of crossover, mutation, and population update
are repeated by a predefined number of iterations, denoted by $I_{\max, 1}$.
In the end, the fittest index set of $\mP_1$, which has the lowest cost function of \eqref{eq:cost},
will be selected as an optimized subsampling index set $\Omega_{\rm opt}$. 
 
Algorithm 1 describes the entire steps of the first-stage GA
to optimize a subsampling index set. 

\begin{table}[!t]
	\fontsize{8}{10pt}\selectfont
	\centering
	\begin{tabular}{l}
		\hlineB{2.5}
		\textbf{Algorithm 1} Genetic Algorithm for Subsampling Optimization \\
		\hline
		\textbf{Input:} Unitary matrix $\Ubu$, Number of measurements $M$, \\ 
		        \qquad \quad Population size $T_1$, Crossover rate $\beta_1$, Mutation rate $\mu_{1}$,  \\
   		        \qquad \quad Maximum number of iterations $I_{\max, 1}$. \\
		\emph{Initialization}: Create a population $\mP_1 = \{ \Omega_1, \cdots, \Omega_{T_1}\}$  \\
		\qquad \qquad \qquad of randomly selected index sets, where $|\Omega_t| = M$.  \\		
		\qquad \qquad \qquad Compute the cost function \eqref{eq:cost} 
		for each index set of $\mP_1$. \\ 
		\emph{Iteration}:  \\
		\textbf{for} $i=1$ to $I_{\max, 1}$ \textbf{do}  \\
		\quad \emph{Crossover}: Create a new population $\mC_1$ with index sets from $\mP_1$. \\
		\quad \emph{Mutation}: Create a new population $\mM_1$ with index sets from $\mP_1$. \\
		\quad \emph{Population update}: Compute the cost function \eqref{eq:cost} for each index set \\ 
		\qquad \qquad \qquad \qquad \quad of $\mI_1 = \mP_1 + \mC_1+\mM_1$, select the $T_1$ index sets \\
		\qquad \qquad \qquad \qquad \quad from $\mI_1$ with the $T_1$ lowest cost functions, and \\
		\qquad \qquad \qquad \qquad \quad update $\mP_1$ with the $T_1$ fittest index sets. \\
		\textbf{end for} \\
		\emph{Selection}: Select the fittest index set $\Omega_{\rm opt}$ from $\mP_1$. \\ 
		\textbf{Output:} Optimized subsampling index set $\Omega_{\rm opt}$ \\
		\hline
	\end{tabular}
	\label{tb:ga}
\end{table}

\subsection{Stage 2: Masking Sequence Optimization}

\subsubsection{Initialization}
In the second-stage GA, a population consists of $T_2$ masking sequences, i.e.,
$\mP_2 = (\vbu_1, \cdots, \vbu_{T_2})$, where
each element of $\vbu_t = (v_{1, t}, \cdots, v_{M, t})^T$ %
is $ v_{m, t} = e^{j \frac{2 \pi a_{m, t}}{N}}$ for $m = 1, \cdots, M$ and $t = 1, \cdots, T_2$.
Initially, $a_{m, t} $ is randomly taken from $\Z_N$.

\subsubsection{Cost Function}
Note that the partial unitary matrix $\Ubu_{\Omega_{\rm opt}}$
is available 
by the optimized subsampling index set $\Omega_{\rm opt}$ from the first-stage GA.
In the second-stage GA, the cost function for a masking sequence $\vbu_t$ is proposed by
\begin{equation}\label{eq:cost2}
	f_2(\Ubu_{\Omega_{\rm opt}, \vbu_t}) = \frac{1}{|\mK_\delta|}\sum_{\pbu_n \in \mK_\delta} {\rm PAPR} (\pbu_n)  ,
\end{equation}
where $\Ubu_{\Omega_{\rm opt}, \vbu_t} = {\rm diag}(\vbu_t) \cdot \Ubu_{\Omega_{\rm opt}}
= [\pbu_1, \cdots, \pbu_N] \triangleq \Pbu$.
In~\eqref{eq:cost2}, $\mK_\delta$ is a set of columns in $\Pbu$
whose PAPR belong to the top $ \delta  \%$, where $|\mK_\delta| = \left\lfloor \frac{\delta N}{100} \right\rfloor$.
That is, $f_2(\Ubu_{\Omega_{\rm opt}, \vbu_t}) $ is the average of top $\delta \%$ PAPR
of $\Pbu$, which will be minimized 
to enhance the PAPR distribution
of the columns of $\Pbu$.

\subsubsection{Crossover}
As in the first-stage GA,
we consider a pair of sequences $\vbu_{t_1}$ and $\vbu_{t_2}$ from $\mP_2$, $1 \leq t_1 \neq t_2 \leq T_2$, where
$f_2 \left(\Ubu_{\Omega_{\rm opt}, \vbu_{t_1}} \right) < f_2 \left(\Ubu_{\Omega_{\rm opt}, \vbu_{t_2}} \right)$.
Then, the first $d_1 = \lceil \beta_2 \cdot M \rceil$ elements from $\vbu_{t_1}$
and the last $d_2 = M - d_1 $ elements from $\vbu_{t_2}$ are combined to generate a new masking sequence,
where $\beta_2 > 0.5$.
Applying the crossover for every pair of sequences in $\mP_2$,
we have a new population $\mC_2$ of size $ \binom{T_2}{2} = \frac{T_2(T_2-1)}{2}$
at each evolution step.

\subsubsection{Mutation}
For mutation, the second-stage GA 
randomly selects $\mu_2$ elements from each sequence 
in $\mP_2$,
where each element is replaced by a new (random) $N$-ary modulated one. 
We obtain
a new population $\mM_2$ of size $T_2$ by applying the mutation
to all sequences in $\mP_2$.

\subsubsection{Population Update}
Through crossover and mutation,
we obtain 
$\mI_2 = \mP_2 + \mC_2 + \mM_2$,
where $ |\mI_2| = 
\frac{T_2(T_2+3)}{2}$.
From $\mI_2$, we select the $T_2$ sequences 
with the $T_2$ lowest cost functions of \eqref{eq:cost2}.
The population $\mP_2$ is then updated by the $T_2$ fittest sequences
at each evolution step. 

\subsubsection{Iteration and Selection}
In the second-stage GA, 
crossover, mutation, and population update
are repeated by a predefined number of iterations, denoted by $I_{\max, 2}$.
Finally, the fittest sequence of $\mP_2$, which has the lowest cost function of \eqref{eq:cost2},
will be selected as an optimized mask $\vbu_{\rm opt}$.

\begin{table}[!t]
	\fontsize{8}{10pt}\selectfont
	\centering
	\begin{tabular}{l}
		\hlineB{2.5}
		\textbf{Algorithm 2} Genetic Algorithm for Masking Sequence Optimization \\
		\hline
		\textbf{Input:} Partial unitary matrix $\Ubu_{\Omega_{\rm opt}}$, Number of measurements $M$, \\ 
		\qquad \quad Population size $T_2$, Crossover rate $\beta_2$, Mutation rate $\mu_{2}$,  \\
		\qquad \quad Maximum number of iterations $I_{\max, 2}$. \\
		\emph{Initialization}: Create a population $\mP_2 = \{ \vbu_1, \cdots, \vbu_{T_2}\}$    \\
		\qquad \qquad \qquad of random modulated $N$-ary sequences of length $M$.  \\		
		\qquad \qquad \qquad Compute the cost function \eqref{eq:cost2} 
		for each sequence of $\mP_2$. \\ 
		\emph{Iteration}:  \\
		\textbf{for} $i=1$ to $I_{\max, 2}$ \textbf{do}  \\
		\quad \emph{Crossover}: Create a new population $\mC_2$ with sequences from $\mP_2$. \\
		\quad \emph{Mutation}: Create a new population $\mM_2$ with sequences from $\mP_2$. \\
		\quad \emph{Population update}: Compute the cost function \eqref{eq:cost2} for each sequence \\ 
		\qquad \qquad \qquad \qquad \quad of $\mI_2 = \mP_2 + \mC_2+\mM_2$, select the $T_2$ sequences \\
		\qquad \qquad \qquad \qquad \quad from $\mI_2$ with the $T_2$ lowest cost functions, and \\
		\qquad \qquad \qquad \qquad \quad update $\mP_2$ with the $T_2$ fittest sequences. \\
		\textbf{end for} \\
		\emph{Selection}: Select the fittest masking sequence $\vbu_{\rm opt}$ from $\mP_2$. \\ 
		\textbf{Output:} Optimized masking sequence $\vbu_{\rm opt}$ \\
		\hline
	\end{tabular}
	\label{tb:ga2}
\end{table}

\begin{figure}
	\centering
	\includegraphics[width=0.46\textwidth, angle=0]{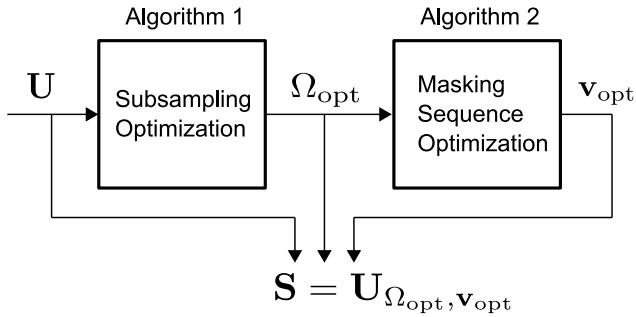}
	\caption{Two-stage GA for non-orthogonal sequence design.}
	\label{fig:twoGA}
\end{figure}

Algorithm 2 describes the entire steps of the second-stage GA
to optimize a masking sequence.
Finally, a set of non-orthogonal sequences, or $\Sbu = \Ubu_{\Omega_{\rm opt}, \vbu_{\rm opt}}$, 
is provided by our two-stage GA of Algorithms 1 and 2,
as illustrated by Fig.~\ref{fig:twoGA}.

\begin{rem}\label{rm:gen}
	The set of non-orthogonal sequences designed by our two-stage GA
	can be represented by 
	\begin{equation}\label{eq:set}
	\Sbu =  \Ubu_{\Omega_{\rm opt}, \vbu_{\rm opt}} = \frac{1}{\sqrt{M}} {\rm diag} (\vbu_{\rm opt}) 
	\cdot \Rbu_{\Omega_{\rm opt}}  \Ubu.
 	\end{equation}	
	Given a unitary matrix $\Ubu$, \eqref{eq:set} means that the matrix
	$\Sbu$ can be generated by the operations of row selection specified by $\Omega_{\rm opt}$ and
	masking by $\vbu_{\rm opt}$.
	Therefore, a BS receiver can generate the sequence set $\Sbu$ easily with 
	the highly structured unitary matrix $\Ubu$ by storing $\Omega_{\rm opt}$ and $\vbu_{\rm opt}$.
	Moreover, 
	CS-based detection can be carried out fast and efficiently at BS, exploiting the fast unitary transform by $\Ubu$.  
	Also, each mMTC device is able to generate its unique sequence on-the-fly 
	with a unique column structure of $\Ubu$ by
	storing $\Omega_{\rm opt}$ and $\vbu_{\rm opt}$, 
	which allows its cheap and efficient implementation. 	
\end{rem}

\section{Simulation Results}

In this section, 
we first demonstrate the effectiveness of our two-stage GA for non-orthogonal sequence design.
Then, we present simulation results of CS-based detection for uplink grant-free access,
which demonstrates the performance of non-orthogonal sequences designed by our two-stage GA.

For the unitary matrix $\Ubu$, 
we use the $N \times N$ Fourier matrix $\Fbu = [F_{k,l}]$,
where
$F_{k,l} = e^{-\frac{j 2 \pi (k-1)(l-1)}{N}}$ 
for $1 \leq k, l \leq N$.
Additionally, we consider another unitary matrix based on Zadoff-Chu (ZC) sequences~\cite{Chu:ZC}.
Each cyclic shift of the ZC sequence of even length 
becomes a column of a matrix $\Zbu = [Z_{k,l}]$,
where each element is given by
\[
Z_{k, l} = e^{\frac{-j \pi (k+N-l)^2}{N} }, \quad 1 \leq k, l \leq N.
\]
Due to the perfect auto-correlation~\cite{Chu:ZC} of ZC sequences,
it is clear that
$\Zbu$, called the ZC matrix, is also unitary.

Beginning with $\Fbu$ and $\Zbu$, 
our two-stage GA gives
$\Fbu_{\Omega_{\rm opt}, \vbu_{\rm opt}}$ and $\Zbu_{\Omega_{\rm opt}, \vbu_{\rm opt}}$, 
respectively, by Algorithms 1 and 2. 
Finally, their columns are
proposed as non-orthogonal sequences, called 
\emph{Fourier-based} and \emph{ZC-based}
sequences, respectively.


\subsection{Effectiveness of Two-Stage GA}
In simulations, 
each stage GA has the population size of $T_1 = T_2 = 20$,
the crossover rate $\beta_1 = \beta_2 = 0.7$, and the mutation number $\mu_1 = \mu_2 = 1$, respectively.
Algorithm 1 has $I_{\max, 1} = 500$,
whereas $I_{\max, 2} = 2000$ in Algorithm 2, as experiments showed that the cost function of Algorithm 2
converges slowly. 
Finally, the cost function of Algorithm 2 computes 
top $\delta = 30 \%$ average of PAPR of $\Pbu= \Fbu_{\Omega_{\rm opt}, \vbu_t} $
(or $\Zbu_{\Omega_{\rm opt}, \vbu_t} $)
for a mask sequence $\vbu_t$.

\begin{figure}[!t]
	\centering
	\includegraphics[width=0.475\textwidth, angle=0]{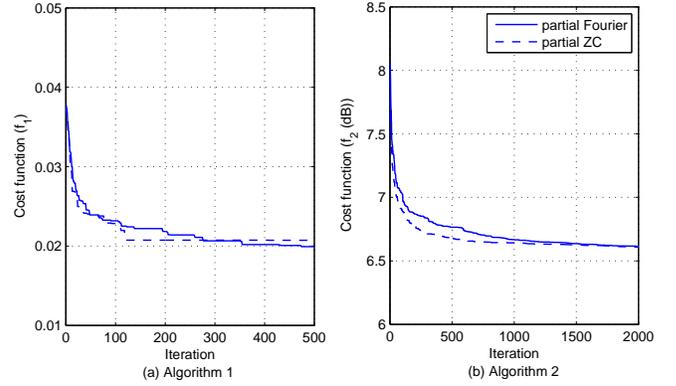}
	\caption{Evolution of the cost functions of Algorithms 1 and 2, where $N = 256$ and $M = 80$. 
		Each curve shows the cost function of the fittest chromosome at each iteration.}
	\label{fig:cost_iter}
\end{figure}

Fig.~\ref{fig:cost_iter} displays the evolution of the cost functions \eqref{eq:cost} and \eqref{eq:cost2} of the fittest chromosomes,
respectively, from $M \times N$ partial Fourier and ZC matrices,
where $N=256$ and $M = 80$.
The figure shows that each stage GA continues to reduce its cost function over the evolution steps.
As mentioned above, we observed that 
the cost function of Algorithm 2 
converges slowly, 
compared to that of Algorithm 1. 
It is because the search space size for the optimization problem \eqref{eq:opt2} is $N^M$,
which is much larger than that of \eqref{eq:opt}, or $\binom{N}{M} \leq \left( \frac{eN}{M} \right)^M$.
Fig.~\ref{fig:cost_iter} also reveals that the cost functions of partial Fourier and ZC matrices
converge to similar values,
which suggests that 
their performance of CS reconstruction and PAPR property will be similar to each other.

\begin{figure}[!h]
	\centering
	\includegraphics[width=0.475\textwidth, angle=0]{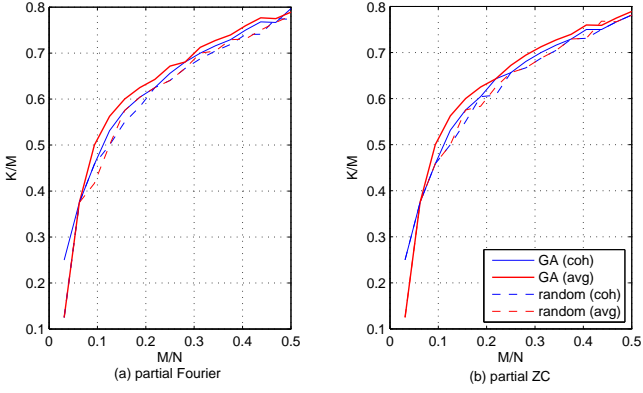}
	\caption{Phase transitions for partial Fourier 
		and ZC 
		matrices under MMV reconstruction (SOMP), 
		where $N=256$ and $J=8$ at ${\rm SNR}= 20$ dB.}
	\label{fig:all_mmv}
\end{figure}

To investigate the effectiveness of the first-stage GA,
we sketch the phase transition diagrams for 
CS reconstruction with partial Fourier and ZC matrices, respectively,
obtained by Algorithm 1. 
We consider an MMV problem $\Ybu = \Abu \Xbu + \Wbu$,
where $ \Abu =  \Fbu_{\Omega_{\rm opt}}$ (or $ \Zbu_{\Omega_{\rm opt}}$),
$\Xbu = (\xbu_1, \cdots, \xbu_J)$ is a jointly sparse matrix 
with common nonzero rows, and 
$\Ybu = (\ybu_1, \cdots, \ybu_J)$ is a collection of $J=8$ measurement vectors. 
The nonzero entries of $\Xbu$ are independently drawn from $\mathcal{CN} (0, 1)$,
where their row positions are uniformly distributed.
Also, each element of $\Wbu$ is the i.i.d. Gaussian noise from $\mathcal{CN} (0, \sigma_n ^2)$,
where the signal-to-noise ratio (SNR) is set to 
${\rm SNR} = \frac{\sum_{t=1} ^J|| \Abu \xbu_t ||_2 ^2}{J K M \sigma_n ^2} = 20$ dB.
In phase transition, we made $10^4$ trials of CS reconstruction
at each test point,
where the step sizes of $\frac{M}{N}$ and $\frac{K}{M}$ are $2^{-5}$ and $10^{-2}$, respectively.
The phase transition
indicates that the corresponding CS reconstruction is successful with 
probability exceeding $99 \%$ below the transition curve, where
a success is declared 
if an estimated $\widehat{\Xbu}$
achieves $\frac{|| \Xbu - \widehat{\Xbu}||_F ^2}{|| \Xbu||_F ^2} < 10^{-2}$.

Fig.~\ref{fig:all_mmv} depicts the phase transitions
for partial Fourier and ZC matrices 
under MMV reconstruction by
the simultaneous orthogonal matching pursuit (SOMP)~\cite{Tropp:somp}, where 
the number of nonzero rows of $\Xbu$ is assumed to be known in advance.
In the figure, 
`GA (avg)' indicates the phase transition of $\Fbu_{\Omega_{\rm opt}}$ (or $\Zbu_{\Omega_{\rm opt}}$)
for which the cost function \eqref{eq:cost} has been minimized by Algorithm 1.
Meanwhile, `GA (coh)' corresponds to the case in which Algorithm 1 
changed its cost function with the mutual coherence, 
i.e., $f_1(\Ubu_{\Omega_t}) 
= \max_{ 1 \leq k \neq l \leq N  }
\frac{\left| \left \langle \abu_{k} ,  \abu_{l}  \right \rangle \right| }
{\| \abu_{k} \|_2 \| \abu_{l} \|_2}$, 
where $\abu_k$ and $\abu_l$ are the $k$th and the $l$th columns of $\Abu = \Ubu_{\Omega_t}$, respectively,
with $\Ubu = \Fbu$ or $\Zbu$.
Also, `random (coh)' and `random (avg)' show the phase transitions for
randomly subsampled Fourier (or ZC) matrices that 
have the lowest coherence and the lowest cost function \eqref{eq:cost}, respectively,
out of $500$ trials.
Fig.~\ref{fig:all_mmv} shows that 
the phase transition curves of `GA (avg)'
are higher than or equal to all the other ones over most compression ratios,
which demonstrates that the partial Fourier and ZC matrices optimized by Algorithm 1 
with the cost function \eqref{eq:cost}
present reliable MMV reconstruction 
over a wide range of compression $\left( \frac{M}{N} \right)$ and sparsity $\left( \frac{K}{M} \right)$ ratios.

\begin{figure}[!t]
	\centering
	\includegraphics[width=0.475\textwidth, angle=0]{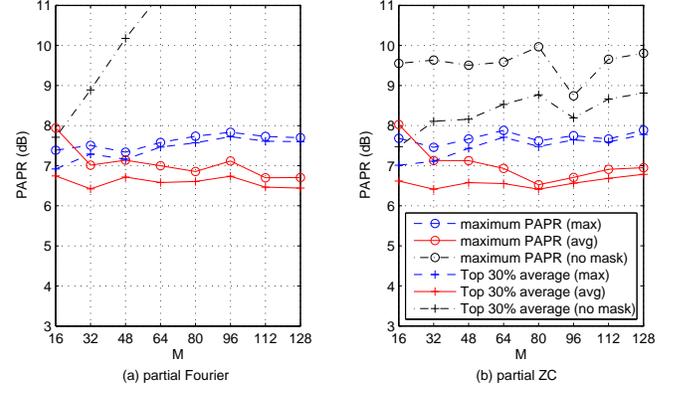}
	\caption{Maximum and top $30 \%$ average 
		PAPR of non-orthogonal sequences from partial Fourier and 
		ZC matrices, 
		where $N=256$.}
	\label{fig:all_papr}
\end{figure}

The effectiveness of the second-stage GA is verified by 
Fig.~\ref{fig:all_papr}, which sketches the maximum and top $ 30 \%$ average PAPR of the sequences
obtained by Algorithm 2. 
In the figure, `avg' means that Algorithm 2 utilized 
the cost function of \eqref{eq:cost2},
while `max' indicates that the maximum PAPR of $\Ubu_{\Omega_{\rm opt}, \vbu_t}$
has been used as the cost function of Algorithm 2, where $\Ubu = \Fbu$ or $\Zbu$.
To demonstrate the PAPR improvement by Algorithm 2, 
we also sketch `no mask', which indicates the PAPR properties 
of the sequences obtained by Algorithm 1 only, or column sequences of $\Fbu_{\Omega_{\rm opt}}$
(or $\Zbu_{\Omega_{\rm opt}}$).
Note that the maximum PAPR of partial Fourier matrices 
for `no mask' is 
outside the scope of this figure, taking the highest value of $M$ 
due to a column of all ones in $\Fbu_{\Omega_{\rm opt}}$.
Fig.~\ref{fig:all_papr} demonstrates that Algorithm 2 can significantly 
reduce the 
maximum and top $30 \%$ average PAPR of 
the sequences from $\Fbu_{\Omega_{\rm opt}}$ and 
$\Zbu_{\Omega_{\rm opt}}$, respectively. 
Also, it 
reveals that using the cost function \eqref{eq:cost2} 
is more effective for Algorithm 2 to enhance the 
PAPR properties. 

\subsection{Performance of CS-based Detection}

Numerical experiments examine the performance of the proposed non-orthogonal sequences
for CS-based AUD and CE in uplink grant-free access.
Under the system model of Section II,
we assume that there are $N = 500$ devices in an mMTC cell,
where each one is assigned a unique non-orthogonal pilot sequence of length $M=80$. 
Reflecting sparse activity, each device sends its pilot with probability $p_a=0.1$
at each access time.
At BS, 
the received signal-to-noise ratio (SNR) per device is set to 
${\rm SNR} = \frac{1}{K} \cdot \frac{\sum_{t=1} ^J|| \Sbu \xbu^{(t)} ||_2 ^2}{J M \sigma_n ^2}$.

For CS-based AUD and CE, 
a BS receiver deploys the SOMP algorithm that requires
no prior knowledge of the number of active devices\footnote{
In simulations, this sparsity-blind SOMP stops its iteration empirically 
if the maximum signal proxy is less than $\sqrt{3\sigma_n ^2 J}$.}.
In AUD, both undetected and false-alarmed devices
are treated as errors.
Thus, the activity error rate (AER) is defined by the average of 
$\frac{|\mS \setminus \widehat{\mS} | + |\widehat{\mS} \setminus \mS  |}{|\mS \cup \widehat{\mS}|}$,
where $\mS$ and $\widehat{\mS}$ are true and detected sets of active devices, respectively.
Also, channel estimation errors are measured by the normalized 
mean squared errors (NMSE), or the average of
$\frac{|| \hbu_\mS - \widehat{\hbu}_{\mS} ||_2 ^2}{|| \hbu_\mS ||_2 ^2}$,
where $\hbu_\mS$ and $ \widehat{\hbu}_{\mS}$ are true and estimated channel vectors,
respectively, for truly active devices.
In simulations, the averages for AER and NMSE are computed over $10^4$ access trials.

To obtain the Fourier- and ZC-based sequences 
from 
$\Fbu_{\Omega_{\rm opt}, \vbu_{\rm opt}}$ and 
$\Zbu_{\Omega_{\rm opt}, \vbu_{\rm opt}}$, 
respectively,
Algorithms 1 and 2 use the same parameters as in Section V.A, but $I_{\max, 1} = 1000$ and $I_{\max, 2} = 4000$.
For comparison, 
we generate complex-valued random Gaussian sequences of length $M$, 
where each element is drawn from the i.i.d. complex Gaussian distribution 
with zero mean and variance $1/M$~\cite{Liu:mimo}.
Also, we use the complex-valued MUSA spreading sequences of length $M$,
where each element is randomly taken from the 3-level signal constellation, i.e.,
$\frac{1}{\sqrt{12}} [{\pm 1} {\pm j} , \pm 1, \pm j, 0]$, in Fig.~2(b) of~\cite{Yuan:MUSA}.
Generating $N$ random Gaussian and $N$ 
MUSA sequences, we have
$M \times N$ matrices $\Gbu$ and $\Mbu$, respectively, 
where each one is a matrix with the lowest coherence 
among $1000$ trials.

The last sequence set for comparison 
is obtained by cyclic shifts of the Zadoff-Chu (ZC) sequences with multiple roots,
where the sequence length $M_{\rm ZC}=79$ is a prime number closest to $M$.
In specific, we begin with an 
$M_{\rm ZC} \times M_{\rm ZC}$ matrix $\Cbu_u$
that consists of all cyclic shifts of 
a $u$th root ZC sequence~\cite{Chu:ZC} of length $M_{\rm ZC}$ with the $k$th element of 
$e^{ \frac{j \pi u k(k+1)}{M_{\rm ZC}} }$, where $u$
is a root number between $1$ and $ M_{\rm ZC}-1$.
Due to the perfect autocorrelation of the ZC sequence, 
$\Cbu_u$ is unitary for any $u$.
For a set of sequences with low PAPR,
we then sort the root numbers $ u=1, \cdots, M_{\rm ZC}-1$
in ascending order of the maximum PAPR
that the column sequences of $\Cbu_u$ achieve.
Taking the first $L = \lceil \frac{N}{M_{\rm ZC}} \rceil$ root numbers, denoted by $u_1, \cdots, u_L$,
we produce a matrix $\Cbu' =  [\Cbu_{u_1}, \cdots, \Cbu_{u_L} ]$,
where the first $N$ columns are finally selected for
an $M_{\rm ZC} \times N$ matrix $\Cbu$, or 
a set of ZC sequences of prime length $M_{\rm ZC}$.
The coherence of $\Cbu$ is $\frac{1}{\sqrt{M_{\rm ZC}}}$, close to the Welch bound equality, 
due to the cross-correlation of ZC sequences with distinct roots~\cite{Sarwate:bound}. 
In simulations, 
the matrix $\Sbu$ in \eqref{eq:mmv} is determined by
the sequence sets under consideration, i.e.,
$\Sbu = \Fbu_{\Omega_{\rm opt}, \vbu_{\rm opt}}, \Zbu_{\Omega_{\rm opt}, \vbu_{\rm opt}}, 
\Gbu, \Mbu$, and $\Cbu$, respectively.


\begin{figure}[!t]
	\centering
	\includegraphics[width=0.365\textwidth, angle=0]{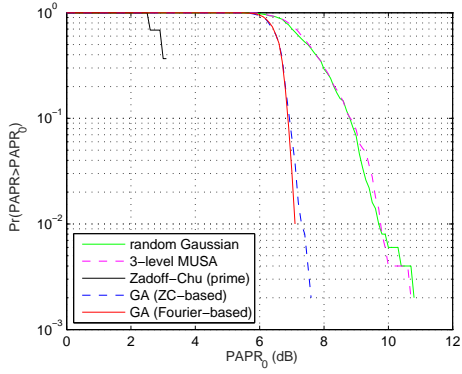}
	\caption{PAPR distribution (CCDF) of non-orthogonal sequences of length $M=80$ ($M_{ZC} = 79$), where $N=500$.}
	\label{fig:all_pd}
\end{figure}

Fig.~\ref{fig:all_pd} displays the complementary cumulative distribution function (CCDF) of 
PAPR of non-orthogonal sequences under consideration, 
where $N=500$ and $M=80~(M_{\rm ZC} = 79)$.
The ZC sequences of prime length $M_{\rm ZC}$, whose PAPR has been optimized as mentioned above,
exhibit the best PAPR distribution with maximum of $3.14 $ dB.
On the other hand, complex-valued random Gaussian and MUSA sequences have the poor distributions,
where the maximum PAPR are $10.86$ dB and $10.79$ dB, respectively.
It is shown that the PAPR distributions of Fourier- and ZC-based sequences from our two-stage GA
are not so good as that of ZC sequences of prime length, but much better than those of the random sequences,
showing the maximum of $7.18$ dB and $7.66$ dB, respectively.
As a result, the PAPR distributions of the proposed sequences appear to be acceptable 
for multicarrier transmission. 

\begin{figure}[!t]
	\centering
	\includegraphics[width=0.475\textwidth, angle=0]{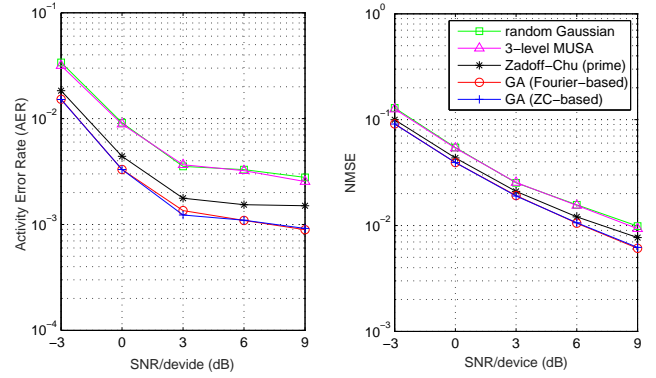}
	\caption{Performance of CS-based AUD and CE of non-orthogonal sequences over the received SNR per device, 
		where $N=500$, $M=80$ ($M_{ZC} = 79$), $J=16$, and $p_a = 0.1$.}
	\label{fig:all_snr}
\end{figure}

\begin{figure}[!t]
	\centering
	\includegraphics[width=0.475\textwidth, angle=0]{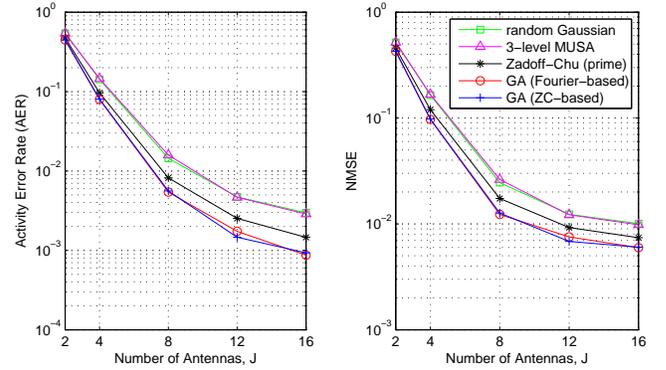}
	\caption{Performance of CS-based AUD and CE of non-orthogonal sequences over the number of antennas, 
		where $N=500$, $M=80$ ($M_{ZC} = 79$), ${\rm SNR} = 9$ dB per device, and $p_a = 0.1$.}
	\label{fig:all_J}
\end{figure}

\begin{figure}[!t]
	\centering
	\includegraphics[width=0.475\textwidth, angle=0]{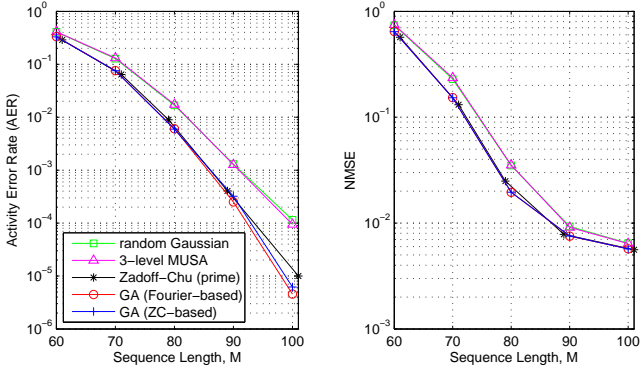}
	\caption{Performance of CS-based AUD and CE of non-orthogonal sequences over the sequence length, 
		where $N=500$, $J=8$, ${\rm SNR} = 5$ dB per device, and $p_a = 0.1$. The prime lengths of ZC sequences are 
		$M_{\rm ZC} = 61, 71, 79, 89,$ and $101$, respectively.}
	\label{fig:all_M}
\end{figure}

Fig.~\ref{fig:all_snr} shows the performance of CS-based AUD and CE 
over the received SNR per device.
In the figure, the AER and NMSE of Fourier- and ZC-based sequences 
from our two-stage GA
are significantly lower than those of complex-valued random Gaussian and MUSA sequences.
The figure also shows that the proposed sequences slightly outperform the ZC sequences of prime length. 
In addition, Figs.~\ref{fig:all_J} and \ref{fig:all_M} depict the AER and NMSE over the number of BS antennas
and the sequence length, respectively,
which also confirm the excellent performance of Fourier- and ZC-based sequences.
Taking into account the difference between $M$ and $M_{\rm ZC}$, 
we can say that 
the AUD and CE performance of the proposed sequences are similar to
those of the ZC sequences of prime lengths.

\subsection{Discussion}
The simulation results of this section demonstrated that the Fourier- and ZC-based sequences
designed by our two-stage GA outperform complex-valued random Gaussian and MUSA sequences
for CS-based AUD and CE. 
Also, we observed that the performance of the proposed sequences is similar to that of   
the ZC sequences of prime lengths. 

In comparison to the ZC sequences, we would like to point out other potential
benefits of the proposed sequences.
First, the non-orthogonal sequences from our two-stage GA
have no limit to the granularity of sequence length, 
which suggests that 
one can obtain the sequences of arbitrary length, 
according to 
the availability of resource subcarriers.
Meanwhile, the length of the ZC sequences should be odd prime only,
which is less flexible for managing resource subcarriers. 
Being able to take arbitrary sequence lengths,
the proposed sequences are expected to
facilitate mMTC systems to manage the resources more effectively for grant-free access.

Second, the proposed sequences constitute a partial unitary matrix with a column mask
through arbitrary row selection,
whereas the ZC sequences of prime length form a deterministic matrix with low coherence.
While the coherence-based recovery guarantee of the deterministic matrix 
is limited by its theoretical bottleneck~\cite{Eldar:CS},
our partial unitary matrix can present theoretical recovery guarantee with higher sparsity,
which ensures reliable CS-based detection theoretically for more active devices in mMTC.
In summary, the proposed non-orthogonal sequences 
can be a good option for uplink grant-free access,
supporting any number of subcarriers and 
providing theoretically guaranteed performance for CS-based AUD and CE.
Remarkably, our GA-based design offers a new set of non-orthogonal sequences
with many advantages over the ZC sequences of prime length,
which are known for the superb performance in practice.

\section{Conclusion}
This paper has presented
a two-stage genetic algorithm (GA) to design new non-orthogonal sequences
for uplink grant-free access in mMTC. 
The first-stage GA is to find a subsampling index set 
for a partial unitary matrix that can be approximated to an ETF.
The second-stage GA then tries to find a masking sequence to be commonly applied  
to each column of the partial unitary matrix from the first-stage,
in order to enhance the PAPR property of the resulting columns.
In each stage GA, a new cost function has been elaborately proposed 
to improve the optimized result.
Finally, the masked columns of the partial unitary matrix
are proposed as new non-orthogonal sequences for uplink grant-free access.
To the best of our knowledge, this is the first effort to apply the GA technique
to non-orthogonal sequence design for achieving low correlation 
and low PAPR properties simultaneously.

Simulation results demonstrated that the
partial Fourier and ZC matrices optimized by our first-stage GA guarantee reliable CS reconstruction
over a wide range of compression and sparsity ratios.
Also, we observed that the second-stage GA produces
the Fourier- and ZC-based sequences that have acceptable PAPR distributions
for multicarrier transmission.
Finally, we demonstrated that 
the Fourier- and ZC-based sequences exhibit reliable performance of CS-based AUD and CE in uplink grant-free access,
which can be suitable for massive connectivity.

The main benefits of this GA-based design are summarized as follows.

\begin{itemize}
	\item[$\bullet$] The non-orthogonal sequences obtained by this GA-based design
	present theoretical recovery guarantee for CS reconstruction by 
	forming a partial unitary matrix through arbitrary row selection.  
	Simulation results confirmed that the Fourier- and ZC-based sequences from this design
	show excellent performance
	of CS-based AUD and CE in uplink grant-free access. 
	\item[$\bullet$] This GA-based design is able to generate non-orthogonal sequences of arbitrary length,
	which can be a good choice for sequence lengths for which algebraically designed sequences with low correlation are unknown.
	In practice, the sequences of arbitrary length can be useful
	for mMTC systems to manage the resources effectively.
	\item[$\bullet$] Based on unitary matrices, this GA-based design
	can offer unimodular sequences of finite phase with rich structure, which 
	are suitable for cost efficient implementation in mMTC devices.	
\end{itemize}

While this GA-based design successfully presented new 
non-orthogonal sequences  
for grant-free massive connectivity,
a further study will be necessary to enhance the design method.
First, the PAPR of the Fourier- and ZC-based sequences, although improved through evolution,
still needs to be reduced further.
It will be challenging, but necessary to devise 
a more elaborate method for further PAPR reduction
so that the PAPR distribution of the resulting sequences can be as good as that of the ZC sequences of prime lengths.
Second, this GA-based design using the Fourier or ZC matrices increases the number of phases of 
sequence elements as more sequences are required to support more mMTC devices.
To resolve this issue, we have employed the Hadamard matrices for our two-stage GA,
but found that the performance of CS-based detection is worse than that of Fourier or ZC matrices.
To obtain sequences of small phase,
we may need to study further with another unitary matrix 
of high dimension but with each element of smaller phase. 
Third, the proposed non-orthogonal sequences are for a single-cell massive connectivity,
but a further study will be necessary for this GA-based design 
to provide multiple sets of non-orthogonal sequences 
for multi-cell environments.
Finally, 
our two-stage GA can be considered 
as a component of DL-based sequence design, 
which is our ongoing research work.



\end{document}